\documentstyle[twocolumn,aps,prl,floats,epsf]{revtex}
\begin{document}
\draft
%\tighten
%* comment this line out for ``long'' draft format
\wideabs{

\title{
Observation of Polarization in Bottomonium Production at 
$\sqrt{s}=38.8$ GeV
}

\author{
C.N.~Brown$^c$, 
T.C.~Awes$^i$,
M.E.~Beddo$^h$, 
M.L.~Brooks$^f$,
J.D.~Bush$^a$,
T.A.~Carey$^f$, 
T.H.~Chang$^h$\cite{byline1},
W.E.~Cooper$^c$,
C.A.~Gagliardi$^j$,
G.T.~Garvey$^f$, 
D.F.~Geesaman$^b$, 
E.A.~Hawker$^{j,f}$, 
X.C.~He$^d$,
L.D.~Isenhower$^a$,
D.M.~Kaplan$^e$, 
S.B.~Kaufman$^b$, 
P.N.~Kirk$^g$, 
D.D.~Koetke$^k$, 
G.~Kyle$^h$,
D.M.~Lee$^f$,
W.M.~Lee$^d$\cite{byline2}, 
M.J.~Leitch$^f$, 
N.~Makins$^b$\cite{byline1}, 
P.L.~McGaughey$^f$, 
J.M.~Moss$^f$,
B.A.~Mueller$^b$,
P.M.~Nord$^k$,
V.~Papavassiliou$^h$, 
B.K.~Park$^f$, 
J.C.~Peng$^f$, 
G.~Petitt$^d$, 
P.E.~Reimer$^{f,b}$,
M.E.~Sadler$^a$,
W.E.~Sondheim$^f$, 
P.W.~Stankus$^i$, 
T.N.~Thompson$^f$, 
R.S.~Towell$^{a,f}$,
R.E.~Tribble$^j$,
M.A.~Vasiliev$^j$\cite{byline3}, 
J.C.~Webb$^h$, 
J.L.~Willis$^a$,
D.K.~Wise$^a$,
%G.R.~Young$^i$\\ 
G.R.~Young$^i$\\ \vspace*{4pt}
(FNAL E866/NuSea Collaboration)\\ \vspace*{2pt}
}
\address{
$^a$Abilene Christian University, Abilene, TX 79699\\
$^b$Argonne National Laboratory, Argonne, IL 60439\\
$^c$Fermi National Accelerator Laboratory, Batavia, IL 60510\\
$^d$Georgia State University, Atlanta, GA 30303\\
$^e$Illinois Institute of Technology, Chicago, IL  60616\\
$^f$Los Alamos National Laboratory, Los Alamos, NM 87545\\
$^g$Louisiana State University, Baton Rouge, LA 70803\\
$^h$New Mexico State University, Las Cruces, NM 88003\\
$^i$Oak Ridge National Laboratory, Oak Ridge, TN 37831\\
$^j$Texas A \& M University, College Station, TX 77843\\
$^k$Valparaiso University, Valparaiso, IN 46383\\
}
\date{\today}

\maketitle
\newpage

\begin{abstract}
We present a measurement of the polarization observed for bottomonium states
produced in
$p$-$Cu$ collisions at $\sqrt{s}=38.8$ GeV. The angular distribution of the
decay dimuons of the $\Upsilon$(1S) state show no polarization at small
$x_F$ and $p_T$ but significant positive
transverse production polarization for either $p_T > 1.8$ GeV/c 
or for $x_F > 0.35$.  
The $\Upsilon$(2S+3S) (unresolved) states 
show a large transverse production polarization
at all values of $x_F$ and $p_T$ measured.
These observations are compared with an NRQCD calculation that predicts a 
transverse polarization in bottomonium production arising from  
quark-antiquark fusion and gluon-gluon fusion diagrams.
\end{abstract}


%* comment this line out for ``long'' draft format
} % end wideabs

%\narrowtext
\vspace{-22pt}
%\medskip

It has been known for some time that the observed production rates of 
charmonium and bottomonium
resonances in hadronic collisions are much larger than the predictions of
lowest order Perturbative Quantum-Chromodynamics (PQCD)~\cite{Bra1998}. 
A calculational approach based upon Non-Relativistic 
Quantum Chromodynamics (NRQCD)
has emerged as a reliable framework for calculating onium 
production~\cite{Bod1}.

Data on the direct production of the charmonium mesons
$\psi$(1S) and $\psi$(2S) at high energies, when compared
with the predictions of NRQCD, indicate that 
color octet contributions dominate the cross section and that S 
state charmonia are produced through gluon fragmentation 
into a $^3S_1^{(8)}$ octet state~\cite{Bra6Cho7}.
Recent investigations have shown that the contribution of color octet states 
to onium production may also be very important at fixed target 
energies, but quantitatively 
the picture is far from complete~\cite{Gup8Ben9}.  In particular, 
NRQCD predictions disagree with measurements of the polarization 
of $\psi$(1S) and $\psi$(2S) mesons produced at collider~\cite{CDF} and fixed 
target energies~\cite{Ting}.

In NRQCD, the predicted spin effects in onium production can provide
further tests of and constraints on the various color octet contributions.
The quark-antiquark fusion and gluon-gluon fusion diagrams which are
expected to dominate onium production at fixed target energies
yield significant transverse polarization~\cite{Khar1} for the 
produced bottomonium mesons
$\Upsilon$(1S), $\Upsilon$(2S), and $\Upsilon$(3S).  The polarization 
results in a $1+\alpha\cos^2(\theta)$ decay angle distribution 
for the polar angle of the decay dimuons
in the Collins-Soper frame~\cite{CollinsSoper}.  Transversely, 
longitudinally, and unpolarized states decay 
with $\alpha$ = $+1$, $-1$, and $0$ respectively.

We have studied the production of dimuons 
in the collision of $800$ GeV/c protons with a copper
beam dump, 
\begin{center}\[p~+~Cu~\rightarrow~\mu^+\mu^-~+~X.\]
\end{center}  
\vspace{-22pt}
The apparatus was originally constructed for Experiment
605~\cite{Crittenden_Jaffe} and was located in the Meson East Laboratory
at Fermilab. The data reported here were taken as part of a subsequent
experiment, Experiment 866~\cite{Hawker}. 
Details of the apparatus used in E866 and a full description of a similar
study of the polarization of dimuons from charmonium states
can be found in Reference~\cite{Ting}.

Here we present polarizations derived from the
angular distribution of 2 million dimuons in the range 
$8.1 < m_{\mu^+\mu^-} < 15.0$ GeV.  The data, after 
analysis cuts, cover the kinematic range $0.0 < x_F < 0.6$ ($x_F$ is the 
fractional longitudinal momentum of the dimuon in 
the nucleon-nucleon center-of-mass frame), and 
$p_T < 4.0$ GeV/c ($p_T$ is the transverse momentum of the dimuon).  

For this measurement the currents of the two spectrometer
magnets were set to 4200 A and 4265 A, their maximum
excitation, which produced a spectrometer acceptance 
that decreased rapidly for 
dimuon masses below $8$ GeV.  Figure 1 shows the 
observed dimuon mass spectrum from $8.1$ GeV to $15.0$ GeV dimuon mass.  
The components of a fit described below are also indicated.
The smooth continuum of dimuons under the bottomonium peaks 
arises from the production of dimuons via quark-antiquark annihilation,
the Drell-Yan process~\cite{DrellYan}.  The experimentally observed 
width of the intrinsically narrow 
onium states arises from muon multiple scattering and energy loss
in the 4m-thick copper target.

\begin{figure}
 \begin{center}
  \mbox{\epsfysize = 3.8 in \epsffile{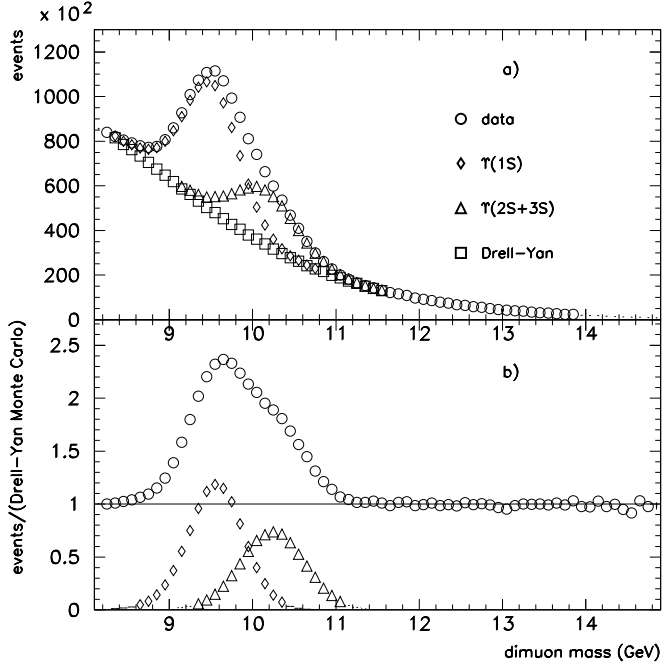}}
   \vspace{-12pt}
 \end{center}
\caption{
a) Mass spectrum of dimuons produced by $800$ GeV 
protons incident on a copper dump. The fit
described in the text for the Drell-Yan dimuon continuum, the $\Upsilon$(1S),
and the sum of the $\Upsilon$(2S) and $\Upsilon$(3S) resonances is also shown.\
b) The ratio of the data to the Monte Carlo-generated Drell-Yan continuum 
events; the ratio of the $\Upsilon$(1S) and the $\Upsilon$(2S+3S) 
generated events to the continuum fit is also shown.}
    \vspace*{-6pt}
  \label{fig1}
\end{figure}

The Drell-Yan dimuon continuum is described well
with a PQCD calculation~\cite{Stirling} incorporating 
a recent MRST determination of the 
proton structure functions~\cite{MRST}.  
The yield of Drell-Yan dimuons is modeled 
with a Monte Carlo simulation of the apparatus 
that generates events as a function of
dimuon $p_T$ and the
apparent fractional momenta, $x_1$ and $x_2$, of the annihilating 
quark-antiquark pair (where $sx_1x_2=m^2$ and $x_1-x_2=x_F$; 
$m$ is the dimuon
mass, $s$ is the center-of-mass energy squared).  A standard 
parametrization of the Drell-Yan production cross section versus $p_T$ was  
fit to the data~\cite{Moreno}.  Drell-Yan virtual photons
are produced transversely polarized and hence their dimuon decay 
is predicted to yield a $1+\cos^2(\theta)$ angular distribution.

Since the mass of a bottomonium state is fixed, the production of a
bottomonium state is a function 
of $p_T$ and $x_F$ only.  The functional
form of the production distributions can be found from the data directly.
Due to
the 330 MeV rms mass resolution of this measurement, we cannot resolve the 
2S and 3S states.  
It has previously been observed that the $p_T$ and $x_F$ distributions
of the $\Upsilon$(2S) and $\Upsilon$(3S) states are very similar~\cite{Moreno}.
Thus, in our fits to the data to extract
the decay angular distributions, we assume that the 2S and 3S states
have the same $p_T$ and $x_F$ distributions.  However, we note that the 
results in this paper are insensitive to this assumption within statistics.

\begin{figure}
 \begin{center}
  \mbox{\epsfysize = 3.8 in \epsffile{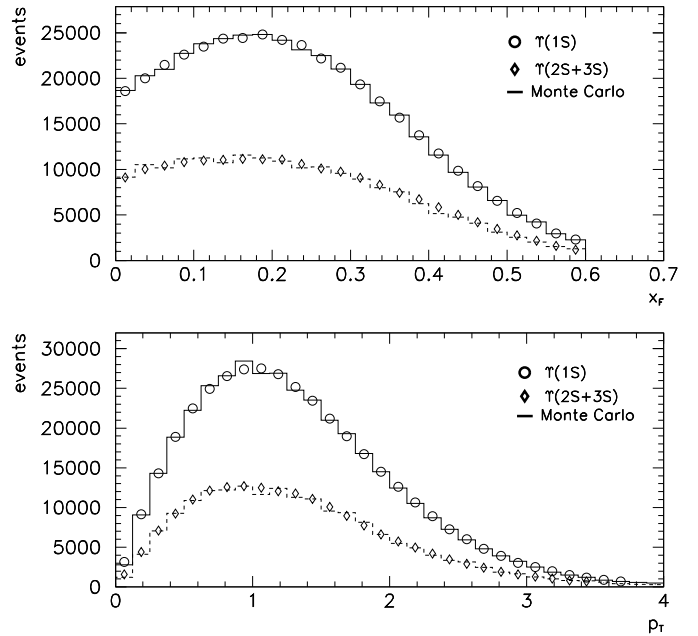}}
   \vspace{-12pt}
 \end{center}
\caption{
The observed $x_F$ and $p_T$ distributions of the $\Upsilon$(1S).
The data distributions are formed by subtracting the 
Monte Carlo-generated Drell-Yan dimuons,  
and the generated $\Upsilon$(2S+3S) dimuon decays, from 
the observed dimuon data.  The Monte Carlo-generated $\Upsilon$(1S) decay
spectra are shown for comparison. A similar comparison is included for the
sum of the $\Upsilon$(2S) and $\Upsilon$(3S) states} 
      \vspace*{-6pt}
   \label{fig2}
\end{figure}

We generated twice 
as many accepted Monte Carlo events as were observed in the data.  
The Drell-Yan dimuon continuum was generated 
using PQCD with MRST parton 
distributions~\cite{MRST}, a shape versus $p_T$ that fit the data, and 
a transverse polarization of 100$\%$.  The Drell-Yan continuum events
were then
weighted with quadratic polynomial functions of $x_1$ and $x_2$ to 
match the data exactly.  The weighting polynomials 
(which varied in value from 0.85 to 1.15) correct for small 
inaccuracies in the modelling of the apparatus and for variations of the p-Cu
cross section from the PQCD prediction (there are known to be small nuclear 
effects in Drell-Yan dimuon yields~\cite{Alde}).
The weighting is important since acceptance correlations between
muon momenta and dimuon decay angle could lead to a false polarization signal
if the observed yield is not modelled correctly versus $x_F$ and $p_T$.

\begin{figure}
 \begin{center}
   \mbox{\epsfysize = 3.8 in \epsffile{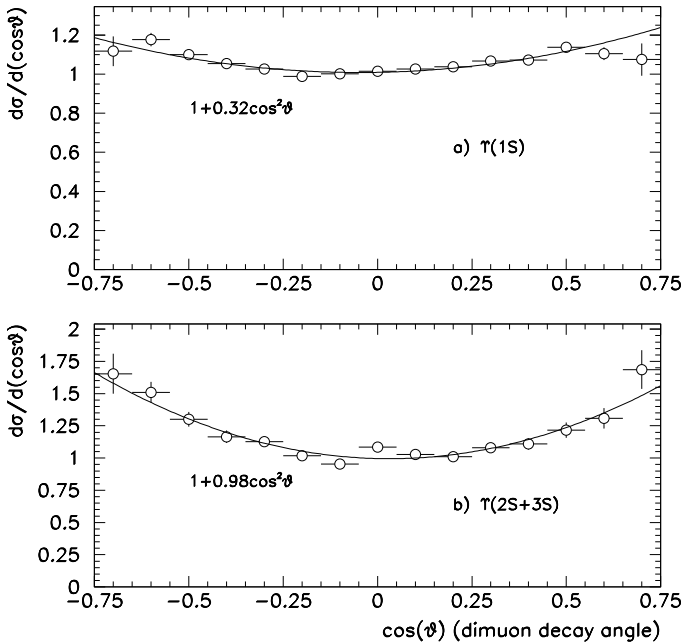}}
    \vspace{-22pt}
 \end{center}
\caption{
a) Decay angular distribution of $\Upsilon$(1S) dimuon decays, formed by 
subtracting the fit contributions of the 
Drell-Yan, $\Upsilon$(2S), and $\Upsilon$(3S) decays from the data (in the bin 
$8.8 < m_{\mu^+\mu^-} < 10.0$ GeV and $p_T > 1.8$ GeV/c). 
A fit to the form
$1+\alpha~cos^2(\theta)$ is superimposed.
b) The corresponding decay distribution for $\Upsilon$(2S+3S) decays 
(for $10.0 < m_{\mu^+\mu^-} < 11.1$ GeV and $p_T > 1.8$ GeV/c). }
    \vspace*{-6pt}
  \label{fig3}
\end{figure}

The Monte Carlo simulation of the bottomonium states 
generated unpolarized $\Upsilon$(1S), $\Upsilon$(2S) and 
$\Upsilon$(3S) events with  
$p_T$ and $x_F$ shapes and a relative 1S/2S/3S weight that matched the data.  
In the final fit to the data, the polarization parameter $\alpha$ of both the 
Drell-Yan continuum and the bottomonium resonances was 
allowed to vary.  The polarizations of the
$\Upsilon$(2S) and $\Upsilon$(3S) 
states were set equal in the fit after attempts to assign different
polarizations to these two states led to large, negatively correlated
statistical errors on $\alpha$ for the two states 
(consistent with the limited resolution mentioned above).

The final fit independently varied the 
shapes of the production distributions and the 
polarizations of the Drell-Yan, $\Upsilon$(1S) and $\Upsilon$(2S+3S)
generated events to match the data. The shapes of
the production distributions agree, within errors,
with those obtained earlier~\cite{Moreno}.
Figure 1a shows the results of the fit
versus dimuon mass for all the data.  Figure 1b shows 
the ratio of the data and 
generated resonances to the generated Drell-Yan events.  
The separation of the
$\Upsilon$(1S) from the combined $\Upsilon$(2S) and $\Upsilon$(3S) states 
is sufficient to yield a stable fit.

In figure 2 we show the $x_F$ and $p_T$ distributions observed
for the $\Upsilon$(1S) ($8.8 < m_{\mu^+\mu^-} < 10.0$ GeV) 
data along with the fitted Monte Carlo 
distributions for the $\Upsilon$(1S).  The data spectra are 
obtained by subtracting the 
Monte Carlo fit distributions for the Drell-Yan, $\Upsilon$(2S), and 
$\Upsilon$(3S) dimuons from the data.  
The figure also includes similar curves for 
the sum of the $\Upsilon$(2S) and $\Upsilon$(3S) states. 
The acceptance varies more slowly 
than the observed event yield versus either $x_F$ or $p_T$; the average $x_F$
of either the analysed $\Upsilon$(1S) or $\Upsilon$(2S+3S) 
data is 0.23 and the average $p_T$ is 1.3 GeV/c.
 
Figure 3 shows the angular distributions, in one 
of four $p_T$ bins, for the
$\Upsilon$(1S) decays and for the sum of the $\Upsilon$(2S) and $\Upsilon$(3S)
decays.  Each
point in figure 3a shows the data in a mass 
bin $8.8 < m_{\mu^+\mu^-} < 10.0$ GeV
with the Monte Carlo-generated contributions from Drell-Yan dimuons, 
$\Upsilon$(2S) decays, and $\Upsilon$(3S) decays subtracted away.  
Similarly, figure 3b shows the data in a mass bin from 
$10.0 < m_{\mu^+\mu^-} < 11.1$ GeV
minus the Monte Carlo-generated Drell-Yan and $\Upsilon$(1S)
events.  The expected $1+\alpha\cos^2(\theta)$ decay angle distribution
fits well in both cases.  The $\chi^2$/DF of the fits 
are 0.7 and 1.2 respectively.

The values of $\alpha$ arising from the combined production distribution and
decay angular distribution fit in the Drell-Yan 
sideband and two onium mass regions
for 4 bins in $p_T$ (bin boundaries at
$p_T$ = 0.0, 0.8, 1.3, 1.8 and 4.0 GeV/c) are 
shown in figure 4a.  The results
versus $x_F$ (4 bins, boundaries at 0.0, 0.12, 0.23, 0.35 and 0.6) 
are shown in figure 4b.  The points are plotted at the 
cross-section-weighted average value of the abscissa.  
A systematic error in $\alpha$ of $\pm$ 0.06 should be added
to the values of the onium polarizations in figure 4.  
This was estimated by varying 
the form of the fitting function for the
Drell-Yan continuum dimuons and by varying the 
width of the mass bins used to fit 
the onium resonances and the Drell-Yan continuum.  

The observed polarization of the Drell-Yan 
continuum dimuons is consistent with 100$\%$ 
transverse polarization in all bins and with previous measurements~\cite{McG}.  
The Drell-Yan sidebands have $0.2 < x_1 < 0.8$ and
$0.06 < x_2 < 0.4$, a region where no significant nuclear 
shadowing is observed~\cite{Vas}.
A fit to the Drell-Yan sideband data (for all $x_F$ and $p_T$) 
yields $\alpha = 1.008 \pm 0.016$ with an estimated systematic error of
$\pm 0.020$.

The $\Upsilon$(1S) data show almost no polarization at small 
$x_F$ and $p_T$.  The data show a finite transverse polarization at
either large $p_T$ or at large $x_F$ (there are no significant 
$x_F$ versus $p_T$ production distribution correlations observed 
in the data).  
\begin{figure}
 \begin{center}
   \mbox{\epsfysize = 3.8 in \epsffile{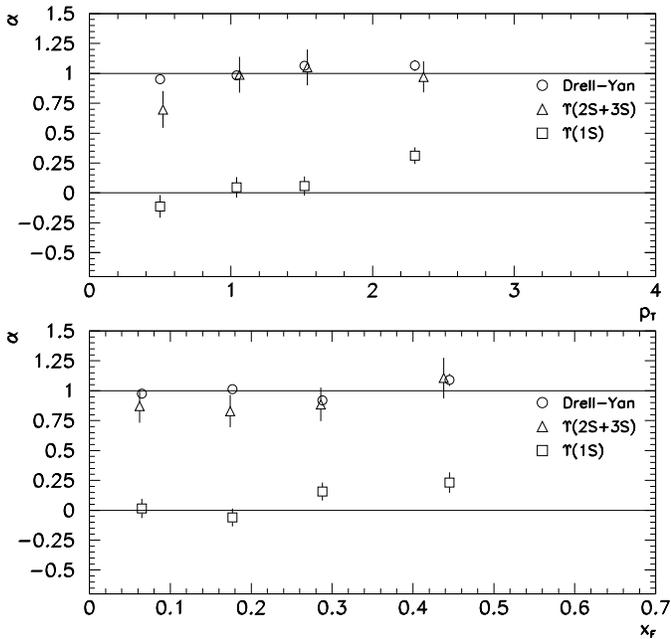}}
    \vspace{-12pt}
 \end{center}
\caption{
a) $\alpha$ versus $p_T$ 
for the Drell-Yan sidebands 
($8.1 < m_{\mu^+\mu^-} < 8.45$ GeV and
$11.1 < m_{\mu^+\mu^-} < 15.0$ GeV), 
$\Upsilon$(1S) ($8.8 < m_{\mu^+\mu^-} < 10.0$ GeV), 
and $\Upsilon$(2S+3S) ($10.0 < m_{\mu^+\mu^-} < 11.1$ GeV).
b) $\alpha$ versus $x_F$ for the same mass regions.  
The errors shown are statistical, there is an additional
systematic error not shown of 0.02 in $\alpha$ for Drell-Yan 
polarizations and 0.06 in $\alpha$ for onium 
polarizations.}
 \vspace*{-6pt}
 \label{fig4}
\end{figure}
This observation disagrees with an 
NRQCD calculation that predicts a polarization of 0.28 to 0.31
at our energies~\cite{Khar1}.  If we fit the 1S state 
for a polarization 
independent of $x_F$ and $p_T$, we get $\alpha = 0.07 \pm 0.04$.

The observation that the polarization of the 
cross-section-weighted average of the 2S+3S states is much larger than 
that of the 1S state at all $x_F$ and $p_T$ contrasts sharply with what is 
seen in the charmonium
system~\cite{CDF}.  Although an NRQCD calculation~\cite{Khar1}
predicts that feeddown decays from higher S, P, and D 
upsilon states dilute the polarization of the 1S state, we can find
no explicit calculation of the polarization expected for the 2S or 3S state.

In the kinematic range $0.0 < x_F < 0.6$ 
and $p_T < 4.0$ GeV/c, the fit to 
the data yields a ratio of $\Upsilon$(2S+3S)/$\Upsilon$(1S)
events of $0.50 \pm 0.01$.  
A separate 3-peak fit yielded an overall ratio of 
$\Upsilon$(3S) to $\Upsilon$(2S) events of $0.46 \pm 0.03$
consistent with previous high resolution measurements~\cite{Moreno}.  
Note that even if the $\Upsilon$(3S) were 100$\%$ polarized, 
the $\Upsilon$(2S) must be at least 35$\%$ polarized to yield the
observed polarizations of the combined peaks. Likewise,  
if the $\Upsilon$(2S) were 100$\%$ polarized, the $\Upsilon$(3S) must have 
significant positive polarization in most bins.

This work was supported in part by the U. S. Department of Energy.

\end{document}